\documentclass[a4paper,11pt]{article}
\usepackage[left=20mm,right=20mm,top=20mm,bottom=20mm]{geometry}
\usepackage{amsmath}
\usepackage{amsfonts}
\usepackage{amssymb}
\usepackage{amsbsy}
\usepackage{color,graphics}
\usepackage{setspace}
\usepackage{epsfig}
\usepackage{cite}
\usepackage{graphicx}
\usepackage{latexsym}
\usepackage{multicol}
\usepackage{footnote}
\renewcommand\textit[1]{{\color{blue}#1}}

\numberwithin{equation}{section}
\begin{document}
\begin{center}
\LARGE{Hyperelastic bodies under homogeneous Cauchy stress induced by three-dimensional non-homogeneous deformations}\\
\vspace{0.3cm}
\large{L. Angela Mihai\footnote{Corresponding author: L. Angela Mihai, Senior Lecturer in Applied Mathematics, School of Mathematics, Cardiff University, Senghennydd Road, Cardiff, CF24 4AG, UK, Email: MihaiLA@cardiff.ac.uk}\quad and\quad Patrizio Neff\footnote{Patrizio Neff, Chair for Nonlinear Analysis and Modelling, Fakult\"{a}t f\"{u}r Mathematik, Universit\"{a}t Duisburg-Essen, Thea-Leymann Sra\ss e 9, 45141 Essen, Germany, Email: patrizio.neff@uni-due.de}}\\
\vspace{0.3cm}

November 5, 2016
\end{center}
\vspace{0.3cm}

\begin{abstract}
In isotropic finite elasticity, unlike in the linear elastic theory, a homogeneous Cauchy stress may be induced by non-homogeneous strains. To illustrate this, we identify compatible non-homogeneous three-dimensional deformations producing a homogeneous Cauchy stress on a cuboid geometry, and provide an example of an isotropic hyperelastic material, which is not rank-one convex, and for which the homogeneous stress and the associated non-homogeneous strains on a domain similar to those analysed are given explicitly. \\


\end{abstract}

\section{Introduction}
In this paper, we consider the question if, and how, a homogeneous Cauchy stress tensor can be generated by non-homogeneous finite deformations. Without assuming invertibility of the Cauchy stress tensor, this question was first addressed in the recent article \cite{Mihai:2016:MN}, where compatible non-homogeneous finite plane deformations with a constant Cauchy stress on a rectangular domain were identified, and an example of an isotropic strain energy function was provided, such that, for a material described by this function and occupying a domain similar to those analysed, the homogeneous stress and the corresponding non-homogeneous strains could be given explicitly. As the energy function in this example was not rank-one convex, it could correspond to materials capable of undergoing phase transitions \cite{Ball:1977b,Ball:1979, Ericksen:1975}.

In isotropic linear elasticity, a homogeneous stress is induced by a homogeneous strain, provided that the usual positive-definiteness assumptions on the elastic energy are assumed. In this case, the linear elastic energy takes the form
\[
W_{\mathrm{lin}}(\nabla\textbf{u})=\mu\ \|\mathrm{dev}\ \mathrm{sym}\nabla\textbf{u}\|^2+\frac{\kappa}{2}\left[\mathrm{tr}\left(\mathrm{sym}\nabla\textbf{u}\right)\right]^2,
\]
where $\textbf{u}: B_{0} \to B$ is the displacement vector, $\boldsymbol{\varepsilon}=\mathrm{sym}\nabla\textbf{u}=\left[\nabla\textbf{u}+(\nabla\textbf{u})^{T}\right]/2$ is the infinitesimal strain tensor, $\mathrm{tr}(\boldsymbol{\varepsilon})=\varepsilon_{11}+\varepsilon_{22}+\varepsilon_{33}$ is the trace of the strain tensor, and
\[
\mathrm{dev}\ \boldsymbol{\varepsilon}=\boldsymbol{\varepsilon}-\frac{1}{3}\mathrm{tr}(\boldsymbol{\varepsilon})\textbf{I}
\]
is the deviatoric strain, with $\textbf{I}$ the tensor identity. In the above formulation, $\|\cdot\|$ denotes to Frobenius norm, hence, for a second order tensor $\textbf{A}$, $\|\textbf{A}\|^2=\textbf{A}:\textbf{A}=\mathrm{tr}(\textbf{A}^T\textbf{A})$.

The corresponding stress-strain law is
\[
\boldsymbol{\sigma}=2\mu\ \mathrm{dev}\ \boldsymbol{\varepsilon}+\kappa\ \mathrm{tr}(\boldsymbol{\varepsilon})\ \textbf{I}.
\]
This relation is invertible if and only if the shear modulus satisfies $\mu> 0$, and similarly the bulk modulus satisfies $\kappa> 0$. 

When $\boldsymbol{\sigma}=\overline{\textbf{T}}$ is given, $\boldsymbol{\varepsilon}=\mathrm{sym}\nabla\textbf{u}=\boldsymbol{\sigma}^{-1}(\overline{\textbf{T}})$ is uniquely determined, and moreover, if $\mathrm{sym}\nabla\textbf{u}=\mathrm{constant}=\boldsymbol{\sigma}^{-1}(\overline{\textbf{T}})\in\mathrm{Sym}(3)$, where $\mathrm{Sym}(3)$ is the set of symmetric matrices, then
\begin{eqnarray}
\nabla\textbf{u}(\textbf{X})&=&\boldsymbol{\sigma}^{-1}(\overline{\textbf{T}})+\textbf{A}(\textbf{X}),\qquad \textbf{A}(\textbf{X})\in\mathfrak{so}(3),\label{Eq:lingrad}
\end{eqnarray}
where $\mathfrak{so}(3)$ is the set of skew-symmetric matrices. This implies
\[
\underbrace{\mathrm{Curl}\ \nabla\textbf{u}}_{=0}=\underbrace{\mathrm{Curl}\ \boldsymbol{\sigma}^{-1}(\overline{\textbf{T}})}_{=0}+\mathrm{Curl}\ \textbf{A}(\textbf{x}),
\]
hence $\mathrm{Curl}\ \textbf{A}(\textbf{X})=0$, and therefore $\textbf{A}(\textbf{X})=\overline{\textbf{A}}=\mathrm{constant}$ \cite{Neff:2008:NM}. 

Thus, a constant stress tensor $\boldsymbol{\sigma}=\overline{\textbf{T}}$ implies the following representation for the displacement field
\begin{eqnarray}
\textbf{u}(\textbf{X})&=&\left[\boldsymbol{\sigma}^{-1}(\overline{\textbf{T}})+\overline{\textbf{A}}\right]\textbf{X}+\overline{\textbf{b}},\label{Eq:linu}
\end{eqnarray}
where $\overline{\textbf{A}}\in \mathfrak{so}(3)$ is arbitrary and $\overline{\textbf{b}}\in\mathrm{R}^3$ is an arbitrary constant translation. Hence, the homogeneous displacement is uniquely determined, up to infinitesimal rigid body rotations and translations, from the constant stress field $\boldsymbol{\sigma}=\overline{\textbf{T}}$.

In nonlinear elasticity, for a homogeneous isotropic hyperelastic body under finite strain deformation, the Cauchy stress tensor can be represented as follows 
\begin{equation}\label{Eq:sigma}
\boldsymbol{\sigma}(\textbf{B})=\beta_{0}\ \textbf{I}+\beta_{1}\ \textbf{B}+\beta_{-1}\ \textbf{B}^{-1},
\end{equation}
where $\textbf{B}=\textbf{F}\textbf{F}^{T}$ is the left Cauchy-Green tensor, with the tensor $\textbf{F}=\nabla\varphi$ representing the deformation gradient, and the coefficients:
\begin{equation}\label{Eq:betas}
\beta_{0}=\frac{2}{\sqrt{I_{3}}}\left(I_{2}\frac{\partial W}{\partial I_{2}}+I_{3}\frac{\partial W}{\partial I_{3}}\right),\qquad
\beta_{1}=\frac{2}{\sqrt{I_{3}}}\frac{\partial W}{\partial I_{1}},\qquad
\beta_{-1}=-2\sqrt{I_{3}}\frac{\partial W}{\partial I_{2}}
\end{equation}
are scalar functions of the strain invariants:
\[
I_{1}(\textbf{B})=\mathrm{tr}\ \textbf{B},\qquad I_{2}(\textbf{B})=\frac{1}{2}\left[\left(\mathrm{tr}\ \textbf{B}\right)^{2}-\mathrm{tr}\ \textbf{B}^{2}\right]=\mathrm{tr}\left(\textrm{Cof}\ \textbf{B}\right),\qquad I_{3}(\textbf{B})=\det\textbf{B},
\]
with $W(I_{1}, I_{2}, I_{3})$ the strain energy density function describing the properties of the isotropic hyperelastic material \cite{Green:1970:GA,Green:1968:GZ,Ogden:1997,TruesdellNoll:2004}.

When the material is incompressible, the Cauchy stress takes the form
\begin{equation}\label{Eq:sigma:inc}
\boldsymbol{\sigma}(\textbf{B})=-p\ \textbf{I}+\beta_{1}\ \textbf{B}+\beta_{-1}\ \textbf{B}^{-1},
\end{equation}
where $p$ is an arbitrary hydrostatic pressure.

If invertibility holds in (\ref{Eq:sigma}), then a unique left Cauchy-Green tensor $\overline{\textbf{B}}\in\mathrm{Sym}^{+}(3)$ can be found that satisfies (see \cite{Ciarlet:1988,Ogden:1997,Ghiba:2015:GNS,Martin:2016:MGN,Neff:2015:NEM,Neff:2015:NGL,Neff:2015:NGLMS})
\begin{eqnarray}
\nabla\varphi\ (\nabla\varphi)^{T}&=&\overline{\textbf{B}}=\boldsymbol{\sigma}^{-1}(\overline{\textbf{T}}).
\end{eqnarray}
This implies 
\begin{equation}
\varphi(\textbf{X})=\left(\overline{\textbf{V}}\ \overline{\textbf{R}}\right)\textbf{X}+\overline{\textbf{b}}=\left[\sqrt{\boldsymbol{\sigma}^{-1}(\overline{\textbf{T}})}\ \overline{\textbf{R}}\right]\textbf{X}+\overline{\textbf{b}},
\end{equation}
where $\overline{\textbf{R}}\in\mathrm{SO}(3)$ is an arbitrary constant rotation, $\overline{\textbf{b}}\in\mathbb{R}^{3}$ is an arbitrary constant translation, and $\overline{\textbf{V}}$ is the left principal stretch tensor satisfying $\overline{\textbf{V}}^2=\overline{\textbf{B}}$, and is uniquely determined from the given stress $\boldsymbol{\sigma}=\overline{\textbf{T}}$ \cite[p. 55]{Ciarlet:1988}.

In this study, we extend the approach developed in \cite{Mihai:2016:MN} to further obtain a homogeneous Cauchy stress generated by non-homogeneous three-dimensional finite deformations. In Section~\ref{sec:function}, we show that the isotropic hyperelastic material characterised by the strain energy function introduced in the context of finite plane deformations can also undergo a homogeneous Cauchy stress that is produced by non-homogeneous three-dimensional deformations. In Section~\ref{sec:geometry}, we construct examples of such deformations on a cuboid domain, where the deformations are continuous and homogeneous in two different parts of the domain, which can only be separated by a plane interface, and the homogeneous deformations corresponding to the two phases are rank-one connected. 

\section{Hyperelastic material under homogeneous stress induced by\\ different deformations}\label{sec:function}

In \cite{Mihai:2016:MN}, the following strain energy function, which is not rank-one convex, was introduced 
\begin{equation}\label{Eq:model:W}
\begin{split}
W&=\frac{\mu}{2}\left(I_{3}^{-1/3}I_{1}-3\right)+\frac{\tilde{\mu}}{4}\left(I_{1}-3\right)^2+\frac{\kappa}{2}\left(I_{3}^{1/2}-1\right)^2\\
&=\frac{\mu}{2}\left[\left\|\frac{\textbf{F}}{\left(\det\textbf{F}\right)^{1/3}}\right\|^2-3\right]+\frac{\tilde{\mu}}{4}\left(\|\textbf{F}\|^2-3\right)^2+\frac{\kappa}{2}\left(\det\textbf{F}-1\right)^2,
\end{split}
\end{equation}
where $\mu>0$ is the infinitesimal shear modulus, $\kappa>0$ is the infinitesimal bulk modulus, and $\tilde{\mu}>0$ is a positive constant independent of the deformation, and $\|\cdot\|$ is the Frobenius norm.

For a material described by the strain energy function (\ref{Eq:model:W}), finite plane elastic deformations were found, such that the corresponding Cauchy stress tensor could be expressed equivalently in terms of two different homogeneous left Cauchy-Green tensors $\textbf{B}=\textbf{F}\textbf{F}^{T}$ and $\widehat{\textbf{B}}=\widehat{\textbf{F}}\widehat{\textbf{F}}^{T}$, and such that some part of the deformed body was under the strain $\textbf{B}$ while another part was under the strain $\widehat{\textbf{B}}$. 

In general, when a Cauchy stress (\ref{Eq:sigma}) can be expressed equivalently in terms of two different homogeneous deformation tensors $\textbf{B}=\textbf{F}\textbf{F}^{T}$ and $\widehat{\textbf{B}}=\widehat{\textbf{F}}\widehat{\textbf{F}}^{T}$, for geometric compatibility, there must exist two non-zero vectors $\textbf{a}$ and $\textbf{n}$, such that the Hadamard jump condition is satisfied as follows \cite{Ball:1987:BJ,Ball:1992:BJ}:
\begin{equation}\label{Eq:rank1an}
\widehat{\textbf{F}}-\textbf{F}=\textbf{a}\otimes\textbf{n},
\end{equation}
where $\textbf{n}$ is the normal vector to the interface between the two phases corresponding to the deformation gradients $\textbf{F}$ and $\widehat{\textbf{F}}$. Equivalently,  $\textbf{F}$ and $\widehat{\textbf{F}}$ must be rank-one connected, \emph{i.e.}
\begin{equation}\label{Eq:rank1}
\mathrm{rank}\left(\textbf{F}-\widehat{\textbf{F}}\right)=1.
\end{equation}

As shown in \cite{Neff:2016:NM}, it is not possible to find a rank-one convex elastic energy, such that the Cauchy stress $\boldsymbol{\sigma}$ is not injective and there exists a homogeneous state with deformation gradient $\textbf{F}$, such that $\sigma(\textbf{F})=\sigma(\textbf{F}+\textbf{a}\otimes\textbf{n})$, with $\textbf{a}$ and $\textbf{n}$ as given in (\ref{Eq:rank1an}).

For the material model (\ref{Eq:model:W}), differentiating with respect to the strain invariants gives:
\begin{eqnarray*}
\frac{\partial W}{\partial I_{1}}&=&\frac{\mu}{2}I_{3}^{-1/3}+\frac{\tilde{\mu}}{2}\left(I_{1}-3\right),\\
\frac{\partial W}{\partial I_{2}}&=&0,\\
\frac{\partial W}{\partial I_{3}}&=&-\frac{\mu}{6}I_{1}I_{3}^{-4/3}+\frac{\kappa}{2}I_{3}^{-1/2}\left(I_{3}^{1/2}-1\right).
\end{eqnarray*}
Hence the material parameters (\ref{Eq:betas}) take the form:
\begin{equation}\label{Eq:betas:example}
\beta_{0}=-\frac{\mu}{3}I_{1}I_{3}^{-5/6}+\kappa\left(I_{3}^{1/2}-1\right),\qquad
\beta_{1}=\mu I_{3}^{-5/6}+\tilde{\mu}I_{3}^{-1/2}\left(I_{1}-3\right),\qquad
\beta_{-1}=0.
\end{equation}

Here, we consider the following two homogeneous deformations with deformation gradients, respectively,
\begin{equation}\label{Eq:FhatF:example}
\textbf{F}=
\left[
\begin{array}{ccc}
k & sa & 0\\
0 & a & 0\\
0 & 0 & 1/a
\end{array}
\right],
\qquad
\widehat{\textbf{F}}=
\left[
\begin{array}{ccc}
k & -sa & 0\\
0 & a & 0\\
0 & 0 & 1/a
\end{array}
\right],
\end{equation}
where $k>0$, $a>0$, and $s>0$ are positive constants. 

For the two deformation gradients given by (\ref{Eq:FhatF:example}), the rank-one connectivity condition (\ref{Eq:rank1}) is satisfied, and is equivalent to
\begin{eqnarray}
\left(F_{11}-\widehat{F}_{11}\right)\left(F_{22}-\widehat{F}_{22}\right)&=&\left(F_{12}-\widehat{F}_{12}\right)\left(F_{21}-\widehat{F}_{21}\right).\label{Eq:model:rank1}
\end{eqnarray}
Note that, when $a=1$, (\ref{Eq:FhatF:example}) correspond to the finite plane deformations analysed in \cite{Mihai:2016:MN}.

The associated left Cauchy-Green tensors are, respectively,
\begin{equation}\label{Eq:BhatB:example}
\textbf{B}=\textbf{F}\textbf{F}^{T}=
\left[
\begin{array}{ccc}
k^2+s^2a^2 & sa^2 & 0\\
sa^2 & a^2 & 0\\
0 & 0 & 1/a^2
\end{array}
\right],
\qquad
\widehat{\textbf{B}}=\widehat{\textbf{F}}\widehat{\textbf{F}}^{T}=
\left[
\begin{array}{ccc}
k^2+s^2a^2 & -sa^2 & 0\\
-sa^2 & a^2 & 0\\
0 & 0 & 1/a^2
\end{array}
\right],
\end{equation}
and, since $s\neq0$, it follows that $\textbf{F}\neq\widehat{\textbf{F}}$ and $\textbf{B}\neq\widehat{\textbf{B}}$. 

By (\ref{Eq:sigma}) and (\ref{Eq:betas:example}), the corresponding Cauchy stresses take the form
\begin{equation}\label{Eq:sigmas:example}
\boldsymbol{\sigma}(\textbf{B})=\beta_{0}\ \textbf{I}+\beta_{1}\ \textbf{B},
\qquad
\boldsymbol{\sigma}(\widehat{\textbf{B}})=\beta_{0}\ \textbf{I}+\beta_{1}\ \widehat{\textbf{B}}.
\end{equation}
Writing the components of the Cauchy stress described by (\ref{Eq:sigmas:example}) in the two equivalent forms $\boldsymbol{\sigma}(\textbf{B})=\boldsymbol{\sigma}(\widehat{\textbf{B}})$ leads to the following simultaneous equations:
\begin{eqnarray}
\beta_{0}+\beta_{1}\ B_{11}&=&\beta_{0}+\beta_{1}\ \widehat{B}_{11},\label{Eq:model:stress1}\\
\beta_{0}+\beta_{1}\ B_{22}&=&\beta_{0}+\beta_{1}\ \widehat{B}_{22}\label{Eq:model:stress2},\\
\beta_{0}+\beta_{1}\ B_{33}&=&\beta_{0}+\beta_{1}\ \widehat{B}_{33}\label{Eq:model:stress3},\\
\beta_{1}\ B_{12}&=&\beta_{1}\ \widehat{B}_{12}\label{Eq:model:stress12},\\
\beta_{1}\ B_{13}&=&\beta_{1}\ \widehat{B}_{13}\label{Eq:model:stress13},\\
\beta_{1}\ B_{23}&=&\beta_{1}\ \widehat{B}_{23}\label{Eq:model:stress23}.
\end{eqnarray}

Since, for both tensors $\textbf{B}$ and $\widehat{\textbf{B}}$ in (\ref{Eq:BhatB:example}), the corresponding invariants are $I_{1}=k^2+s^2a^2+a^2+1/a^2$ and $I_{3}=k^2$, by (\ref{Eq:betas:example}),
\[
\begin{split}
\beta_{0}&=-\frac{\mu}{3}k^{-5/3}\left(k^2+s^2a^2+a^2+\frac{1}{a^2}\right)+\kappa\left(k-1\right),\\
\beta_{1}&=\mu k^{-5/3}+\tilde{\mu}k^{-1}\left(k^2+s^2a^2+a^2+\frac{1}{a^2}-3\right).
\end{split}
\]

In this case, it can be verified that, if 
\[
\frac{\mu}{3\tilde{\mu}}<\left(\frac{3-a^2-1/a^2}{4}\right)^{4/3}
\quad\mbox{and}\quad 
0<s<\frac{1}{a}\sqrt{3-4\left(\frac{\mu}{3\tilde{\mu}}\right)^{3/4}-a^2-\frac{1}{a^2}},
\]
then there exists $k_{0}\in(0,1)$, such that, for $k=k_{0}$,
\[
\beta_{0}=-\frac{\mu}{3}k_{0}^{-5/3}\left(k_{0}^2+s^2a^2+a^2+\frac{1}{a^2}\right)-\kappa\left(1-k_{0}\right)<0,\qquad
\beta_{1}=0,
\]
and, therefore, the equations (\ref{Eq:model:rank1}) and (\ref{Eq:model:stress1})-(\ref{Eq:model:stress23}) are satisfied simultaneously, with the common Cauchy stress tensor 
\begin{equation}\label{Eq:model:stress}
\boldsymbol{\sigma}(\textbf{B})=\boldsymbol{\sigma}(\widehat{\textbf{B}})=\beta_{0}\textbf{I}.
\end{equation}

As in the case of finite plane deformations described in \cite{Mihai:2016:MN}, when $k\to1$, $a\to1$, and $s\to0$, corresponding to the linear elastic limit, in (\ref{Eq:BhatB:example}), $k^2+s^2a^2+a^2+1/a^2-3$ is arbitrarily small, and $\beta_{1}=\mu k^{-5/3}+\tilde{\mu}k^{-1}\left(k^2+s^2a^2+a^2+1/a^2-3\right)\to\mu\neq0$. Hence, if $\beta_{1}=0$ and $s$ is close to zero, then $k$ and $a$ cannot be simultaneously close to one, \emph{i.e.} the two rank-one connected deformation gradients (\ref{Eq:FhatF:example}) do not correspond to infinitesimal deformations.

If different $k_{1},k_{2}\in(0,1)$ exist, such that $\beta_{1}=0$ with the same $s>0$ and $a>0$, then two different sets of deformation gradients
\[
\textbf{F}=
\left[
\begin{array}{ccc}
k_{1} & sa & 0\\
0 & a & 0\\
0 & 0 & 1/a
\end{array}
\right]
\qquad\mbox{and}\qquad
\widehat{\textbf{F}}=
\left[
\begin{array}{ccc}
k_{1} & -sa & 0\\
0 & a & 0\\
0 & 0 & 1/a
\end{array}
\right],
\]
satisfy (\ref{Eq:rank1}) and produce the same Cauchy stress
\[
\boldsymbol{\sigma}=\beta_{0}\ \textbf{I}=\left[-\frac{\mu}{3}k_{1}^{-5/3}\left(k_{1}^2+s^2a^2+a^2+\frac{1}{a^2}\right)-\kappa\left(1-k_{1}\right)\right]\textbf{I}, 
\]
and,
\[
\textbf{F}=
\left[
\begin{array}{ccc}
k_{2} & sa & 0\\
0 & a & 0\\
0 & 0 & 1/a
\end{array}
\right]
\qquad\mbox{and}\qquad
\widehat{\textbf{F}}=
\left[
\begin{array}{ccc}
k_{2} & -sa & 0\\
0 & a & 0\\
0 & 0 & 1/a
\end{array}
\right],
\]
are rank-one connected and produce the Cauchy stress
\[
\boldsymbol{\sigma}=\beta_{0}\ \textbf{I}=\left[-\frac{\mu}{3}k_{2}^{-5/3}\left(k_{2}^2+s^2a^2+a^2+\frac{1}{a^2}\right)-\kappa\left(1-k_{2}\right)\right]\textbf{I}.
\]

Next, we show that, an elastic body made from a material described by the strain energy function (\ref{Eq:model:W}) and occupying a particular three-dimensional domain can undergo a constant Cauchy stress  (\ref{Eq:model:stress}), with some part of the body deforming under the strain $\textbf{B}$ while another part deforms under the strain $\widehat{\textbf{B}}$, where $\textbf{B}$ and $\widehat{\textbf{B}}$ are given by (\ref{Eq:BhatB:example}). 

\section{Homogeneous stress induced by non-homogeneous deformations}\label{sec:geometry}

We consider a continuous material body that occupies a compact three-dimensional domain $\overline\Omega\in\mathbb{R}^{3}$, such that the interior of the body is an open, bounded, connected set $\Omega\subset\mathbb{R}^{3}$, and its boundary $\Gamma=\partial\Omega=\overline\Omega\setminus\Omega$ is Lipschitz continuous (in particular, we assume that a unit normal vector $\textbf{n}$ exists almost everywhere on $\Gamma$). The body is subject to a finite elastic deformation defined by the one-to-one, orientation preserving transformation
\[
\boldsymbol{\varphi}:\Omega\to\mathbb{R}^{3},
\]
such that  $J=\det\left(\nabla\boldsymbol{\varphi}\right)>0$ on $\Omega$  and $\boldsymbol{\varphi}$ is injective on $\Omega$. The injectivity condition on $\Omega$ guarantees that interpenetration of the matter is avoided. However, since self-contact is permitted, this transformation does not need to be injective on $\bar\Omega$.

We denote by $\textbf{x}=\boldsymbol{\varphi}(\textbf{X})$ the spatial point corresponding to the place occupied by the particle $\textbf{X}$ in the deformation $\boldsymbol{\varphi}$. For the deformed body, the equilibrium state in the presence of a dead load is described in terms of the Cauchy stress by the Eulerian field equation
\begin{eqnarray}
-\mathrm{div}\ \boldsymbol{\sigma}(\textbf{x})&=&\textbf{f}(\textbf{x}), \qquad \textbf{x}\in\varphi(\Omega).\label{Eq:equilibrium}
\end{eqnarray}
The governing equation (\ref{Eq:equilibrium}) is completed by a constitutive law for $\boldsymbol{\sigma}$, depending on material properties, and supplemented by boundary conditions.

Since the domain occupied by the body after deformation is generally unknown, we rewrite the above equilibrium problem as an equivalent problem in the reference configuration where the independent variables are $\textbf{X}\in\Omega$. The corresponding Lagrangian equation of nonlinear elastostatics takes the form
\begin{eqnarray}
-\mathrm{Div}\ \textbf{S}_{1}(\textbf{X})&=&\textbf{f}(\textbf{X}), \qquad \textbf{X}\in\Omega,\label{Eq:balance}
\end{eqnarray}
where $\textbf{S}_{1}=\boldsymbol{\sigma}\ \mathrm{Cof}\ \textbf{F}$ is the first Piola-Kirchhoff stress tensor, $\textbf{F}=\nabla\boldsymbol{\varphi}$ is the gradient of the deformation $\boldsymbol{\varphi}(\textbf{X})=\textbf{x}$, such that $J=\det\textbf{F}>0$,  and $\textbf{f}(\textbf{X})=J\ \textbf{f}(\textbf{x})$.

For a homogeneous compressible hyperelastic material described by the strain energy function $W(\textbf{F})$, the first Piola-Kirchhoff stress tensor is equal to
\begin{eqnarray}
\textbf{S}_{1}(\textbf{F})=\frac{\partial W(\textbf{F})}{\partial\textbf{F}},\label{Eq:PK1}
\end{eqnarray}
and the associated Cauchy stress tensor takes the form $\boldsymbol{\sigma}=J^{-1}\textbf{S}_{1}\textbf{F}^{T}=\textbf{S}_{1}\left(\mathrm{Cof}\textbf{F}\right)^{-1}$.

\subsection{The boundary value problem}

The general boundary value problem (BVP) is to find the displacement $\textbf{u}(\textbf{X})=\varphi(\textbf{X})-\textbf{X}$, for all $\textbf{X}\in\Omega$, such that the equilibrium equation (\ref{Eq:balance}) is satisfied subject to the following conditions on the relatively disjoint, open subsets of the boundary $\{\Gamma_{D},\Gamma_{N}\}\subset\partial\Omega$, such that $\partial \Omega\setminus\left(\Gamma_{D}\cup\Gamma_{N}\right)$ has zero area \cite{LeTallec94,Mihai:2013:MG,Oden:1972}:
\begin{itemize}
\item The Dirichlet (displacement) conditions on $\Gamma_{D}$
\begin{eqnarray}
\textbf{u}(\textbf{X})=\textbf{u}_{D}(\textbf{X}),\label{Eq:Dbc}
\end{eqnarray}
\item The Neumann (traction) conditions on $\Gamma_{N}$, 
\begin{eqnarray}
\textbf{S}_{1}(\textbf{X})\textbf{N}=\textbf{g}_{N}(\textbf{X}),\label{Eq:Nbc}
\end{eqnarray}
where $\textbf{N}$ is the outward unit normal vector to $\Gamma_{N}$, and $\textbf{g}_{N}dA=\boldsymbol{\tau}da$, where $\boldsymbol{\tau}=\boldsymbol{\sigma}\textbf{n}$ is the surface traction measured per unit area of the deformed state.
\end{itemize}

The existence of a solution to the BVP depends on whether or not there exists a deformation which minimises, in the local or global sense, the total elastic energy of the body \cite{Ball:1977a,Ball:1977b,Schroeder:2010:SN}. In particular, if the Cauchy stress is constant, then the equilibrium equation (\ref{Eq:equilibrium}) is satisfied.

\subsection{Compatible finite deformations}
\begin{figure}[htbp]
\begin{center}
\includegraphics[width=0.45\textwidth]{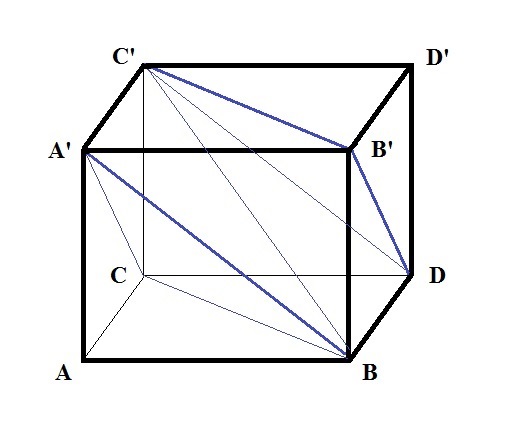}
\caption{Cuboid $ABCDA'B'C'D'$ partitioned into six right-angled tetrahedra: $ A'ABC$, $ A'C'BC$, $ A'B'C'B$, $ B'C'BD$, $ B'C'D'D$, $ C'BCD$. }\label{Fig:abcd}
\end{center}
\end{figure}

We consider the finite deformation of an elastic cuboid partitioned into uniform right-angled tetrahedra as shown in Figure~\ref{Fig:abcd}.

Assuming that the deformation gradient is homogeneous in every tetrahedron, in a single tetrahedron $ A'ABC$, the displacement field takes the form
\begin{equation}\label{SP:eq:app:u}
\textbf{u}(\textbf{X})=\left[
\begin{array}{c}
u_{1}(\textbf{X})\\
u_{2}(\textbf{X})\\
u_{3}(\textbf{X})\\
\end{array}
\right]=
\left[
\begin{array}{c}
a_{11}X_{1}+a_{12}X_{2}+a_{13}X_{3}+b_{1}\\
a_{21}X_{1}+a_{22}X_{2}+a_{23}X_{3}+b_{2}\\
a_{31}X_{1}+a_{32}X_{2}+a_{33}X_{3}+b_{3}
\end{array}
\right],
\end{equation}
where the 12 undetermined coefficients $a_{ij}$ and $b_{i}$, with $i,j=1,2,3$, are constants, and the associated deformation gradient is equal to
\begin{eqnarray}
\textbf{F}&=&\left[
\begin{array}{ccc}
1+a_{11} & a_{12} & a_{13}\\
a_{21} & 1+a_{22} & a_{23}\\
a_{31} & a_{32}   & 1+a_{33}
\end{array}
\right].\label{Eq:F}
\end{eqnarray}

In order to determine the coefficients $a_{ij}$ and $b_{i}$, with $i,j=1,2,3$, we first evaluate the displacement (\ref{SP:eq:app:u}) at the 4 vertices $\{A, B, C, A'\}\in\mathbb{R}^3$:
\[
A'=\left[
\begin{array}{c}
X_{1}^{A'}\\
X_{2}^{A'}\\
X_{3}^{A'}
\end{array}
\right],\qquad
A=\left[
\begin{array}{c}
X_{1}^A\\
X_{2}^A\\
X_{3}^A
\end{array}
\right],\qquad
B=\left[
\begin{array}{c}
X_{1}^B\\
X_{2}^B\\
X_{3}^B
\end{array}
\right],\qquad
C=\left[
\begin{array}{c}
X_{1}^C\\
X_{2}^C\\
X_{3}^C
\end{array}
\right].
\]

The displacements $\left\{\textbf{u}^{A'}, \textbf{u}^{A}, \textbf{u}^{B}, \textbf{u}^{C}\right\}\in\mathbb{R}^3$ at the 4 vertices are, respectively,
\[
\textbf{u}^{A'}=\left[
\begin{array}{c}
u_{1}^{A'}\\
u_{2}^{A'}\\
u_{3}^{A'}
\end{array}
\right],\qquad
\textbf{u}^A=\left[
\begin{array}{c}
u_{1}^A\\
u_{2}^A\\
u_{3}^A
\end{array}
\right],\qquad
\textbf{u}^B=\left[
\begin{array}{c}
u_{1}^B\\
u_{2}^B\\
u_{3}^B
\end{array}
\right],\qquad
\textbf{u}^C=\left[
\begin{array}{c}
u_{1}^C\\
u_{2}^C\\
u_{3}^C
\end{array}
\right].
\]

Hence, a system of 12 linear equations can be formed from which the 12 coefficients are expressed uniquely in terms of the displacements, as follows:
\[
\begin{split}
a_{11}&=
\det\left[
\begin{array}{cccc}
u_{1}^{A'} & X_{2}^{A'} & X_{3}^{A'} & 1\\
u_{1}^A & X_{2}^A & X_{3}^A & 1\\
u_{1}^B & X_{2}^B & X_{3}^B & 1\\
u_{1}^C & X_{2}^C & X_{3}^C & 1
\end{array}
\right]
\det\left[
\begin{array}{cccc}
X_{1}^{A'} & X_{2}^{A'} & X_{3}^{A'} & 1\\
X_{1}^A & X_{2}^A & X_{3}^A & 1\\
X_{1}^B & X_{2}^B & X_{3}^B & 1\\
X_{1}^C & X_{2}^C & X_{3}^C & 1
\end{array}
\right]^{-1},\\
a_{12}&=\det\left[
\begin{array}{cccc}
X_{1}^{A'} & u_{1}^{A'} & X_{3}^{A'} & 1\\
X_{1}^A & u_{1}^A & X_{3}^A & 1\\
X_{1}^B & u_{1}^B & X_{3}^B & 1\\
X_{1}^C & u_{1}^C & X_{3}^C & 1
\end{array}
\right]
\det\left[
\begin{array}{cccc}
X_{1}^{A'} & X_{2}^{A'} & X_{3}^{A'} & 1\\
X_{1}^A & X_{2}^A & X_{3}^A & 1\\
X_{1}^B & X_{2}^B & X_{3}^B & 1\\
X_{1}^C & X_{2}^C & X_{3}^C & 1
\end{array}
\right]^{-1},\\
a_{13}&=\det\left[
\begin{array}{cccc}
X_{1}^{A'} & X_{2}^{A'} & u_{1}^{A'} & 1\\
X_{1}^A & X_{2}^A & u_{1}^A & 1\\
X_{1}^B & X_{2}^B & u_{1}^B & 1\\
X_{1}^C & X_{2}^C & u_{1}^C & 1
\end{array}
\right]
\det\left[
\begin{array}{cccc}
X_{1}^{A'} & X_{2}^{A'} & X_{3}^{A'} & 1\\
X_{1}^A & X_{2}^A & X_{3}^A & 1\\
X_{1}^B & X_{2}^B & X_{3}^B & 1\\
X_{1}^C & X_{2}^C & X_{3}^C & 1
\end{array}
\right]^{-1},
\end{split}
\]
\[
\begin{split}
a_{21}&=
\det\left[
\begin{array}{cccc}
u_{2}^{A'} & X_{2}^{A'} & X_{3}^{A'} & 1\\
u_{2}^A & X_{2}^A & X_{3}^A & 1\\
u_{2}^B & X_{2}^B & X_{3}^B & 1\\
u_{2}^C & X_{2}^C & X_{3}^C & 1
\end{array}
\right]
\det\left[
\begin{array}{cccc}
X_{1}^{A'} & X_{2}^{A'} & X_{3}^{A'} & 1\\
X_{1}^A & X_{2}^A & X_{3}^A & 1\\
X_{1}^B & X_{2}^B & X_{3}^B & 1\\
X_{1}^C & X_{2}^C & X_{3}^C & 1
\end{array}
\right]^{-1},\\
a_{22}&=\det\left[
\begin{array}{cccc}
X_{1}^{A'} & u_{2}^{A'} & X_{3}^{A'} & 1\\
X_{1}^A & u_{2}^A & X_{3}^A & 1\\
X_{1}^B & u_{2}^B & X_{3}^B & 1\\
X_{1}^C & u_{2}^C & X_{3}^C & 1
\end{array}
\right]
\det\left[
\begin{array}{cccc}
X_{1}^{A'} & X_{2}^{A'} & X_{3}^{A'} & 1\\
X_{1}^A & X_{2}^A & X_{3}^A & 1\\
X_{1}^B & X_{2}^B & X_{3}^B & 1\\
X_{1}^C & X_{2}^C & X_{3}^C & 1
\end{array}
\right]^{-1},\\
a_{23}&=\det\left[
\begin{array}{cccc}
X_{1}^{A'} & X_{2}^{A'} & u_{2}^{A'} & 1\\
X_{1}^A & X_{2}^A & u_{2}^A & 1\\
X_{1}^B & X_{2}^B & u_{2}^B & 1\\
X_{1}^C & X_{2}^C & u_{2}^C & 1
\end{array}
\right]
\det\left[
\begin{array}{cccc}
X_{1}^{A'} & X_{2}^{A'} & X_{3}^{A'} & 1\\
X_{1}^A & X_{2}^A & X_{3}^A & 1\\
X_{1}^B & X_{2}^B & X_{3}^B & 1\\
X_{1}^C & X_{2}^C & X_{3}^C & 1
\end{array}
\right]^{-1},
\end{split}
\]
\[
\begin{split}
a_{31}&=
\det\left[
\begin{array}{cccc}
u_{3}^{A'} & X_{2}^{A'} & X_{3}^{A'} & 1\\
u_{3}^A & X_{2}^A & X_{3}^A & 1\\
u_{3}^B & X_{2}^B & X_{3}^B & 1\\
u_{3}^C & X_{2}^C & X_{3}^C & 1
\end{array}
\right]
\det\left[
\begin{array}{cccc}
X_{1}^{A'} & X_{2}^{A'} & X_{3}^{A'} & 1\\
X_{1}^A & X_{2}^A & X_{3}^A & 1\\
X_{1}^B & X_{2}^B & X_{3}^B & 1\\
X_{1}^C & X_{2}^C & X_{3}^C & 1
\end{array}
\right]^{-1},\\
a_{32}&=\det\left[
\begin{array}{cccc}
X_{1}^{A'} & u_{3}^{A'} & X_{3}^{A'} & 1\\
X_{1}^A & u_{3}^A & X_{3}^A & 1\\
X_{1}^B & u_{3}^B & X_{3}^B & 1\\
X_{1}^C & u_{3}^C & X_{3}^C & 1
\end{array}
\right]
\det\left[
\begin{array}{cccc}
X_{1}^{A'} & X_{2}^{A'} & X_{3}^{A'} & 1\\
X_{1}^A & X_{2}^A & X_{3}^A & 1\\
X_{1}^B & X_{2}^B & X_{3}^B & 1\\
X_{1}^C & X_{2}^C & X_{3}^C & 1
\end{array}
\right]^{-1},\\
a_{33}&=\det\left[
\begin{array}{cccc}
X_{1}^{A'} & X_{2}^{A'} & u_{3}^{A'} & 1\\
X_{1}^A & X_{2}^A & u_{3}^A & 1\\
X_{1}^B & X_{2}^B & u_{3}^B & 1\\
X_{1}^C & X_{2}^C & u_{3}^C & 1
\end{array}
\right]
\det\left[
\begin{array}{cccc}
X_{1}^{A'} & X_{2}^{A'} & X_{3}^{A'} & 1\\
X_{1}^A & X_{2}^A & X_{3}^A & 1\\
X_{1}^B & X_{2}^B & X_{3}^B & 1\\
X_{1}^C & X_{2}^C & X_{3}^C & 1
\end{array}
\right]^{-1},
\end{split}
\]
\[
\begin{split}
b_{1}&=
\det\left[
\begin{array}{cccc}
X_{1}^{A'} & X_{2}^{A'} & X_{3}^{A'} & u_{1}^{A'}\\
X_{1}^A & X_{2}^A & X_{3}^A & u_{1}^A\\
X_{1}^B & X_{2}^B & X_{3}^B & u_{1}^B\\
X_{1}^C & X_{2}^C & X_{3}^C & u_{1}^C
\end{array}
\right]
\det\left[
\begin{array}{cccc}
X_{1}^{A'} & X_{2}^{A'} & X_{3}^{A'} & 1\\
X_{1}^A & X_{2}^A & X_{3}^A & 1\\
X_{1}^B & X_{2}^B & X_{3}^B & 1\\
X_{1}^C & X_{2}^C & X_{3}^C & 1
\end{array}
\right]^{-1},\\
b_{2}&=\det\left[
\begin{array}{cccc}
X_{1}^{A'} & X_{2}^{A'} & X_{3}^{A'} & u_{2}^{A'}\\
X_{1}^A & X_{2}^A & X_{3}^A & u_{2}^A\\
X_{1}^B & X_{2}^B & X_{3}^B & u_{2}^B\\
X_{1}^C & X_{2}^C & X_{3}^C & u_{2}^C
\end{array}
\right]
\det\left[
\begin{array}{cccc}
X_{1}^{A'} & X_{2}^{A'} & X_{3}^{A'} & 1\\
X_{1}^A & X_{2}^A & X_{3}^A & 1\\
X_{1}^B & X_{2}^B & X_{3}^B & 1\\
X_{1}^C & X_{2}^C & X_{3}^C & 1
\end{array}
\right]^{-1},\\
b_{3}&=\det\left[
\begin{array}{cccc}
X_{1}^{A'} & X_{2}^{A'} & X_{1}^{A'} & u_{3}^{A'}\\
X_{1}^A & X_{2}^A & X_{1}^A & u_{3}^A\\
X_{1}^B & X_{2}^B & X_{1}^B & u_{3}^B\\
X_{1}^C & X_{2}^C & X_{1}^C & u_{3}^C
\end{array}
\right]
\det\left[
\begin{array}{cccc}
X_{1}^{A'} & X_{2}^{A'} & X_{3}^{A'} & 1\\
X_{1}^A & X_{2}^A & X_{3}^A & 1\\
X_{1}^B & X_{2}^B & X_{3}^B & 1\\
X_{1}^C & X_{2}^C & X_{3}^C & 1
\end{array}
\right]^{-1}.
\end{split}
\]

First, we demonstrate that it is possible for an elastic body occupying a cuboid domain to deform such that the deformation gradient is equal to $\textbf{F}$ on some part of the body and to $\widehat{\textbf{F}}\neq\textbf{F}$ on another part.

For every tetrahedron, there are $12$ unknown coefficients of the form $a_{ij}$ and $b_{i}$, with $i,j=1,2,3$, hence $12m^2$ such coefficients for the entire domain, which can be determined uniquely in terms of the displacements as discussed above. 

It remains to find the equations from which the displacements are computed. Given the continuity of the displacement fields at the $(m+1)^3$ vertices, there are $3$ displacement components $u_{i}$, with $i=1,2,3$, for every vertex, \emph{i.e.} $3(m+1)^3$ displacement components in total. After the boundary conditions are imposed, $6(m-1)^2+12(m-1)+8$ systems of 3 algebraic equations each, \emph{i.e.} $18(m-1)^2+36(m-1)+24$ algebraic equations in total are provided at the vertices situated on the boundary. This leaves $3(m+1)^3-18(m-1)^2-36(m-1)-24=3(m-1)^3$ displacement components, corresponding to the interior vertices, for which additional information is needed. This information may come, for example, from the condition that, on each tetrahedron which does not have a vertex on the boundary, the determinant of the deformation gradient is equal to some given positive constant $d$, which is always valid for incompressible materials where $d=1$, and which can generate the required remaining equations. 

Hence, the displacement fields, which are continuous at the vertices, and the corresponding deformation gradients, which may differ from one tetrahedron to another, can be uniquely determined from the boundary conditions and the constraint that the determinant of the deformation gradient is a given constant. In this case, the rank-one connectivity of the deformation gradients would mean additional constraints on the solution, and must be taken into account a priori, \emph{i.e.} when imposing the boundary conditions. 

Note that, even though the displacements are continuous at each vertex, the deformation gradient, and hence the left Cauchy-Green tensor, may differ from one tetrahedron to another. Indeed, on a different right-angled tetrahedron $ A'C'BC$, the displacement field takes the form
\begin{equation}\label{SP:eq:app:uhat}
\widehat{\textbf{u}}(\textbf{X})=\left[
\begin{array}{c}
\widehat{u}_{1}(\textbf{X})\\
\widehat{u}_{2}(\textbf{X})\\
\widehat{u}_{3}(\textbf{X})\\
\end{array}
\right]=
\left[
\begin{array}{c}
\widehat{a}_{11}X_{1}+\widehat{a}_{12}X_{2}+\widehat{a}_{13}X_{3}+\widehat{b}_{1}\\
\widehat{a}_{21}X_{1}+\widehat{a}_{22}X_{2}+\widehat{a}_{23}X_{3}+\widehat{b}_{2}\\
\widehat{a}_{31}X_{1}+\widehat{a}_{32}X_{2}+\widehat{a}_{33}X_{3}+\widehat{b}_{3}
\end{array}
\right],
\end{equation}
and the coefficients $\widehat{a}_{ij}$ and $\widehat{b}_{i}$, with $i,j=1,2,3$, can be uniquely computed from the displacements $\left\{\widehat{\textbf{u}}^{A'}, \widehat{\textbf{u}}^{C'}, \widehat{\textbf{u}}^{B}, \widehat{\textbf{u}}^{C}\right\}\in\mathbb{R}^3$ at the 4 vertices $\{A', C', B, C\}\in\mathbb{R}^3$, respectively. The corresponding deformation gradient is 
\begin{eqnarray}
\widehat{\textbf{F}}&=&\left[
\begin{array}{ccc}
1+\widehat{a}_{11} & \widehat{a}_{12} & \widehat{a}_{13}\\
\widehat{a}_{21} & 1+\widehat{a}_{22} & \widehat{a}_{23}\\
\widehat{a}_{31} & \widehat{a}_{23}   & 1+\widehat{a}_{33}
\end{array}
\right].\label{Eq:Fhat}
\end{eqnarray}

Given that the displacements are continuous at the vertices $A'$, $B$, $C$ which are common to both tetrahedra $ A'ABC$ and $ A'C'BC$, \emph{i.e.} $\textbf{u}^{A'}=\widehat{\textbf{u}}^{A'}$, $\textbf{u}^{B}=\widehat{\textbf{u}}^{B}$, $\textbf{u}^{C}=\widehat{\textbf{u}}^{C}$, the following 3 systems of 3 algebraic equations each are obtained:
\begin{eqnarray*}
a_{11}X_{1}^{A'}+a_{12}X_{2}^{A'}+a_{13}X_{3}^{A'}+b_{1}&=&\widehat{a}_{11}X_{1}^{A'}+\widehat{a}_{12}X_{2}^{A'}+\widehat{a}_{13}X_{2}^{A'}+\widehat{b}_{1},\\
a_{21}X_{1}^{A'}+a_{22}X_{2}^{A'}+a_{23}X_{3}^{A'}+b_{2}&=&\widehat{a}_{21}X_{1}^{A'}+\widehat{a}_{22}X_{2}^{A'}+\widehat{a}_{23}X_{3}^{A'}+\widehat{b}_{2},\\
a_{31}X_{1}^{A'}+a_{32}X_{2}^{A'}+a_{33}X_{3}^{A'}+b_{3}&=&\widehat{a}_{31}X_{1}^{A'}+\widehat{a}_{32}X_{2}^{A'}+\widehat{a}_{33}X_{3}^{A'}+\widehat{b}_{3},
\end{eqnarray*}
\begin{eqnarray*}
a_{11}X_{1}^{B}+a_{12}X_{2}^{B}+a_{13}X_{3}^{B}+b_{1}&=&\widehat{a}_{11}X_{1}^{B}+\widehat{a}_{12}X_{2}^{B}+\widehat{a}_{13}X_{2}^{B}+\widehat{b}_{1},\\
a_{21}X_{1}^{B}+a_{22}X_{2}^{B}+a_{23}X_{3}^{B}+b_{2}&=&\widehat{a}_{21}X_{1}^{B}+\widehat{a}_{22}X_{2}^{B}+\widehat{a}_{23}X_{3}^{B}+\widehat{b}_{2},\\
a_{31}X_{1}^{B}+a_{32}X_{2}^{B}+a_{33}X_{3}^{B}+b_{3}&=&\widehat{a}_{31}X_{1}^{B}+\widehat{a}_{32}X_{2}^{B}+\widehat{a}_{33}X_{3}^{B}+\widehat{b}_{3},
\end{eqnarray*}
\begin{eqnarray*}
a_{11}X_{1}^{C}+a_{12}X_{2}^{C}+a_{13}X_{3}^{C}+b_{1}&=&\widehat{a}_{11}X_{1}^{C}+\widehat{a}_{12}X_{2}^{C}+\widehat{a}_{13}X_{2}^{C}+\widehat{b}_{1},\\
a_{21}X_{1}^{C}+a_{22}X_{2}^{C}+a_{23}X_{3}^{C}+b_{2}&=&\widehat{a}_{21}X_{1}^{C}+\widehat{a}_{22}X_{2}^{C}+\widehat{a}_{23}X_{3}^{C}+\widehat{b}_{2},\\
a_{31}X_{1}^{C}+a_{32}X_{2}^{C}+a_{33}X_{3}^{C}+b_{3}&=&\widehat{a}_{31}X_{1}^{C}+\widehat{a}_{32}X_{2}^{C}+\widehat{a}_{33}X_{3}^{C}+\widehat{b}_{3},
\end{eqnarray*}
from which the 6 free coefficients $b_{i}$ and $\widehat{b}_{i}$, with $i=1,2,3$, as well as the 3 coefficients $\widehat{a}_{3j}$, with $j=1,2,3$, can be expressed in terms of the remaining coefficients $a_{ij}$ and $\widehat{a}_{ij}$. Hence, the deformation gradient $\textbf{F}$ in $ A'ABC$ may differ from the deformation gradient $\widehat{\textbf{F}}$ in $ A'C'BC$.

\begin{figure}[htbp]
\begin{center}
\includegraphics[width=0.45\textwidth]{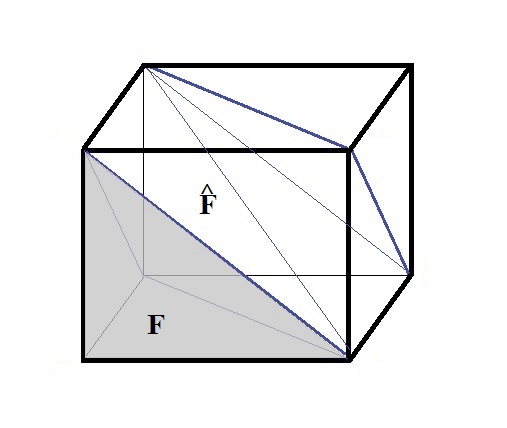}\qquad
\includegraphics[width=0.5\textwidth]{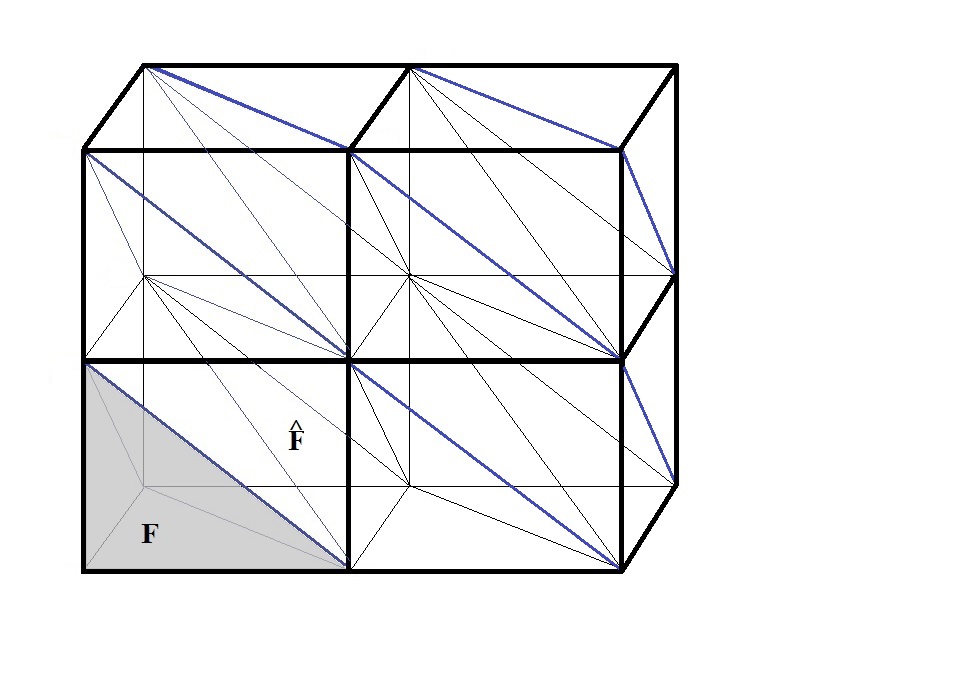}
\caption{Examples of cuboids partitioned into right-angled tetrahedra, showing the rank-one connected deformation gradients $\textbf{F}$ and $\widehat{\textbf{F}}$ in two parts of the body, which can only be separated by a plane surface.}\label{Fig:abcd-fhat}
\end{center}
\end{figure}

Next, we show that, for a cuboid domain partitioned into right-angled tetrahedra, if the deformation is continuous throughout the domain and the deformations gradient is equal to $\textbf{F}$ on one set of tetrahedra and to $\widehat{\textbf{F}}$ on the remaining set, then the common faces between the two sets must lie in the same plane.

Assuming that there are 4 common vertices $[X^{(k)}_{1},X^{(k)}_{2}]^{T}$, with $k=1,2,3,4$, which are not co-planar, at each of these vertices, the displacements are continuous, \emph{i.e.} the following 12 identities hold simultaneously:
\begin{eqnarray*}
a_{11}X^{(k)}_{1}+a_{12}X^{(k)}_{2}+a_{13}X^{(k)}_{3}+b_{1}&=&\widehat{a}_{11}X^{(k)}_{1}+\widehat{a}_{12}X^{(k)}_{2}+\widehat{a}_{13}X^{(k)}_{3}+\widehat{b}_{1},\\
a_{21}X^{(k)}_{1}+a_{22}X^{(k)}_{2}+a_{23}X^{(k)}_{3}+b_{2}&=&\widehat{a}_{21}X^{(k)}_{1}+\widehat{a}_{22}X^{(k)}_{2}+\widehat{a}_{23}X^{(k)}_{3}+\widehat{b}_{2},\\
a_{31}X^{(k)}_{1}+a_{32}X^{(k)}_{2}+a_{33}X^{(k)}_{3}+b_{3}&=&\widehat{a}_{31}X^{(k)}_{1}+\widehat{a}_{32}X^{(k)}_{2}+\widehat{a}_{33}X^{(k)}_{3}+\widehat{b}_{3},
\end{eqnarray*}
with $k=1,2,3,4$, which implies $a_{ij}=\widehat{a}_{ij}$ and $b_{i}=\widehat{b}_{i}$, with $i,j=1,2,3$, \emph{i.e.} $\textbf{F}=\widehat{\textbf{F}}$.

Thus, if the deformation is continuous throughout the deforming body, and the deformation gradient is $\textbf{F}$ on one set of right-angled tetrahedra and $\widehat{\textbf{F}}$ on the remaining set, then a plane surface must separate the two sets (see Figure~\ref{Fig:abcd-fhat}). Consequently, there are no layers of the domain where these sets can alternate.

\section{Conclusion}
As established in \cite{Mihai:2016:MN}, for finite elastic deformations, a homogeneous Cauchy stress is not always induced by homogeneous strains. Here, we showed that the isotropic hyperelastic material described by the strain energy function introduced by \cite{Mihai:2016:MN} in the context of finite plane deformations could further undergo a homogeneous Cauchy stress generated by three-dimensional non-homogeneous deformations, and identified examples of such deformations on a cuboid geometry. Since the energy function in our example is not rank-one convex, it may correspond to materials capable of undergoing phase transitions.

\paragraph{Acknowledgements.}
The support for L. Angela Mihai by the Engineering and Physical Sciences Research Council of Great Britain under research grant EP/M011992/1 is gratefully acknowledged. 


\end{document}